\documentclass[pra, twocolumn, nofootinbib, superscriptaddress, citeautoscript, nobibnotes, nofootinbib, aps]{revtex4-2}

\usepackage[]{standalone}
\usepackage{import}
\usepackage[utf8]{inputenc}
\usepackage[T1]{fontenc}
\usepackage{cmap}
\usepackage{graphicx}
\usepackage{mathtools}

\usepackage[hyperfootnotes=false,breaklinks=true]{hyperref}
\usepackage{setspace}
\hypersetup{
 bookmarksopen=true,
 bookmarksopenlevel=1,
 colorlinks=true,
 linkcolor=blue,
 anchorcolor=blue,
 citecolor=blue,
 filecolor=blue,
 urlcolor=blue,
 pdfpagemode=UseOutlines,
 pdfstartview={XYZ null null 1},
 linktocpage=true,
}
\usepackage{amsmath}
\allowdisplaybreaks  
\usepackage{amssymb}
\usepackage{amstext}
\usepackage{amsfonts}
\usepackage{mathrsfs}
\usepackage{amsthm}
\usepackage{dsfont}
\usepackage{bm}
\usepackage{braket}
\usepackage{physics}
\usepackage{enumitem}
\usepackage[
 capitalise,
]{cleveref}

\usepackage{pgfplots}
\pgfplotsset{compat=newest}
\usepackage{tikz}
\usetikzlibrary{
  positioning,
  shapes.geometric,
  arrows,
  arrows.meta,
  backgrounds,
  fit,
  intersections,
}
\usepackage[english]{babel}
\usepackage{qcircuit}
\usepackage{algorithm,algorithmic}

\usepackage{comment}

\theoremstyle{plain}

\theoremstyle{definition}

\newcommand{\mycomment}[1]{}

\def\defeq{\mathrel{\mathop:}=}

\usepackage{relsize}
\usepackage{cancel}

\makeatletter
\def\blfootnote{\xdef\@thefnmark{}\@footnotetext}

\begin{document}

\blfootnote{This manuscript has been authored by UT-Battelle, LLC, under Contract No. DE-AC0500OR22725 with the U.S. Department of Energy. The United States Government retains and the publisher, by accepting the article for publication, acknowledges that the United States Government retains a non-exclusive, paid-up, irrevocable, worldwide license to publish or reproduce the published form of this manuscript, or allow others to do so, for the United States Government purposes. The Department of Energy will provide public access to these results of federally sponsored research in accordance with the DOE Public Access Plan.}

\title{Gauge-Fixing Quantum Density Operators At Scale}

\author{Amit Jamadagni}  
\email{gangapurama@ornl.gov}
\affiliation{Computational Sciences and Engineering Division, %
Oak Ridge National Laboratory, %
Oak Ridge, Tennessee 37831, USA}

\author{Eugene Dumitrescu}  
\email{dumitrescuef@ornl.gov}
\affiliation{Computational Sciences and Engineering Division, %
Oak Ridge National Laboratory, %
Oak Ridge, Tennessee 37831, USA}

\begin{abstract}
We provide theory, algorithms, and simulations of non-equilibrium quantum systems using a one-dimensional (1D) completely-positive (CP), matrix-product (MP) density-operator ($\rho$) representation. By generalizing the matrix product state's orthogonality center, to additionally store positive classical mixture correlations, the MP$\rho$ factorization naturally emerges. In this setting, we analytically and numerically examine the virtual gauge freedoms associated with the representation of quantum density operators. Based on this perspective, we simplify algorithms in certain limits to speed up the integration of the canonical form's master equation dynamics. This enables us to quickly evolve under the dynamics of two-body quantum channels without resorting to optimization-based methods. In addition to this technical advance, we also scale-up numerical examples and discuss implications for accurately modeling hardware architectures and predicting their performance in the near term. This includes an example of the quantum to classical transition of informationally leaky, i.e., decohering, qubits. In this setting, due to loss from environmental interactions, non-local complex coherence correlations are converted into global incoherent classical statistical mixture correlations. Lastly, the representation of both global and local correlations is discussed. We expect this work to have applications in additional non-equilibrium settings, beyond qubit engineering.

\end{abstract}
\maketitle

\section{Introduction} \label{sec:Introduction}
The physical principle of locality refers to the notion that bodies interact (and become entangled) locally in space-time\footnote{Long ranged interactions originate from \textit{localized} sources and the resulting forces propagate at, or below, the speed of light}. Locality, in turn, justifies spatio-temporal causality. In addition to its importance for fundamental physics, \textit{information}'s locality holds tremendous implications for efficient computations. Indeed, the cost of transporting data between distant memory locations has become a paradigmatic bottleneck in parallelized and asynchronous post-Moore's-law computations.  

The matrix product state (MPS) rigorously defines the density matrix renormalization group algorithm~\cite{White1992} in terms of a data-structure which maps local data onto a physically localized memory array. Along with new 2D forms~\cite{Zaletel2020}, these data-structures leverage representation-theoretic redundancies (mathematical gauge invariances) to manifestly encode local properties with local data-structures. In these examples, local properties are locally encoded by gauging multi-linear (ML) tensor network (TN) data structures into canonical forms which satisfy mathematical constraints~\cite{Perez-Garcia2007}. Mapping local properties to local data structures aids in the representation, optimization, and computation of local quantities, even with respect to highly-entangled quantum states. In doing so, ML representations (MLReps) minimize the computational resources in terms of the memory allocation and data communication requirements.

In addition to coherence correlations, responsible for quantum interference phenomena, storing incoherent mixture populations complicates potential scalable representations of density operators ($\rho$reps). Prior scalable $\rho$rep efforts have been hindered by positivity breaking numerical instabilities--eventually leading to unphysical probabilities--or overly simple models of states and processes which do not accurately capture physically relevant non-equilibrium and open quantum system dynamics. While a rigorous theory has been presented for pure quantum states \cite{Perez-Garcia2007}, work is required to elevate density operator ML data-structures to the same level of representation-theoretic mathematical rigor. 

In addition to each $\ket{\psi_k}$, a density operator \begin{equation} \label{eq:rho} \rho=\sum_{k=1}^{\kappa} p_k |\psi_k \rangle \langle \psi_k| \end{equation} contains non-negative ($0 \leq p_k \leq 1$) $L_{1,1}$-normalized ($\sum_k p_k = 1$) incoherent correlations $\vec{p} = \{p_1, \cdots, p_{\kappa} \}$.\footnote{Pure states have $\text{dim}[\vec{p}]=p_1=1$; thermal states $\rho_\beta \propto e^{- \beta H}$, at finite temperature $T=(\beta k_B)^{-1}$, have $\text{dim}[\vec{p}] = \text{dim}[H]$} Given these properties, we call $\vec{p}$ an \textit{incoherent, classical probability} distribution. In this article we develop machinery to handle and analyze incoherent correlations. In doing so, we see how different physical processes generate both local and global incoherent distributions. 

In Sec.~\ref{sec:MPS}, we briefly review the relation between complex amplitudes and coherence correlations before defining the MPS in terms of its canonical constraints. To further represent density operators, which contain both quantum and classical correlations, Sec.~\ref{sec:MPρ} reviews a completely positive generalization of the MPS canonical form. In Sec.~\ref{sec:num_sim}, we model processes which convert coherence correlations into mixture correlations. In doing so, we entangle qubits and also integrate the open quantum system dynamics which reveals a flow towards noise-localized, informationally-trivial fixed-points. We also examine the \textit{global} mixture correlations that emerge from quantum erasure and particle loss channels, which have applications to quantum architectures based on photons and atomic traps. Finally, in Sec.~\ref{sec:conclusions} we summarize the key results and further elucidate domains, such as applications to the characterization and engineering of noisy qubits, that can potentially benefit from MP$\rho$'s algorithms.

\section{Matrix Product States} \label{sec:MPS}

In this section, we briefly review the notion of coherence correlations as singular values of a quantum wavefunction and show how this leads to the MPS form. 

\subsection{Quantum correlations and entanglement} 
\label{sec:CC}
Consider a pure state $|\Psi \rangle$ bi-partitioned into left (L) and right (R) regions. The Schmidt decomposition of $|\Psi \rangle$, i.e., singular value decomposition (SVD), factorizing the state into L and R degrees of freedom, reads $|\Psi \rangle  = \sum_\lambda c_\lambda |\psi_\lambda \rangle_L \otimes |\phi_\lambda \rangle_R$. Here $\vec{c}=\{c_1,\cdots,c_{\chi}\}$ (see App.~\ref{app:C_tensor} for more details) is a list of quantum \textit{coherence correlation} amplitudes which are $L_2-$normalized: $\sum_{\lambda=1}^{\lambda =\chi} |c_\lambda|^2 = 1$. When L and R each consist of a single spin-1/2 particle (qubit), the entanglement is maximized by the locus of Bell states which are defined as states with entanglement correlations $\vec{c}=\{c_1,c_2\} = \{ \frac{1}{\sqrt{2}},\frac{1}{\sqrt{2}} \}$. 

If dim$[\vec{c}] = c_1 = 1$, we have $|\Psi \rangle  = |\psi \rangle_L \otimes |\phi \rangle_R$, which means that,  $|\phi \rangle$ and $|\psi \rangle$ are \textit{disentangled} and representable using memory resources describing the \textit{individual} sub-systems. In contrast to the entangled system's multiplicative representational scaling, causing the exponential curse of dimensionality, the memory footprint of disentangled quantum states scales \textit{additively}. A consequence is that the representation and manipulation of disentangled systems is embarrassingly parallelizable. 

Tri-partite qubit states may likewise be represented with a correlation vector $|\Psi \rangle = \sum_{s_i \in \{0,1\}} c_{s_1,s_2,s_3} |s_1\rangle \otimes |s_2 \rangle  \otimes |s_3 \rangle$ where $s_i$ refers to the region $i$'s local 
Hilbert space. Spin-1/2s (qubits) have $s_i \in \{\uparrow, \downarrow\} (\{0, 1\})$ while spin-1 particles (qutrits) have $s_i \in \{\uparrow, 0, \downarrow\} (\{0, 1, 2\})$, and so on for $d$-dimensional systems.

\subsection{MPS Canonical Form}\label{sec:MLCRep}

Instead of \textit{explicitly} storing the coherence correlations, we may \textit{implicitly encode} them via an MLRep. This is indeed the case for the MPS representation~\cite{Verstraete_2008}, where an additional trace over virtual dimensions $\chi$ encodes coherence correlations as 
\begin{equation}
\label{eq:MPS}
    \ket{\Psi} = \sum_{s_i} \text{Tr}_\chi[A^{(1)} \cdots A^{(N)}] |s_1\rangle \otimes \cdots \otimes |s_N\rangle,
\end{equation}
with $s_i$ now representing the $i^{\text{th}}$ qubit's computational basis states. The $A^{(i)}$'s are multi-dimensional tensors obeying
\begin{equation}
A^{(i)}:=A^{\chi_{i},s_i}_{\chi_{i+1}}  = \mathds{V}_{\chi_{i}} \otimes \mathds{V}_{s_i} \otimes \mathds{V}^*_{\chi_{i+1}} \rightarrow \mathds{C}. 
\end{equation}
$\chi$ is called the virtual bond dimension, and it encodes the space of coherent correlations. Equation~\ref{eq:MPS} tells us that the individual amplitudes are found by contracting over the virtual $\chi$ dimensions associated to the each of the $A^{(i)}$ tensors. For the remainder of this work we abuse notation and refer to virtual indices by their dimensionality. Following the notation of Ref.~\onlinecite{SCHOLLWOCK2011}, the grouping of  $\chi_{i}, s_{i}$ as a superscript index and $\chi_{i+1}$ as a subscript reflects a  \textit{left-orthogonalization} with the orthogonality center (OC) at site $i$. Similarly, \textit{right-orthogonalization} groups $s_i$ with $\chi_{i+1}$ instead of with $\chi_i$.

Canonical MPS data-structures are then rigorously defined as those satisfying a pair of quadratic  constraints~\cite{Perez-Garcia2007}:
\begin{subequations}
    \begin{eqnarray}
    \label{eq:constrains}
    &&\sum_i A^{(i)}  {A^{(i)}}^\dagger =  \mathds{1}_{\chi_i}, \; \forall \; 1\leq i \neq \text{OC} \leq N,
    \label{eq:iso} \\
    &&\sum_i A^{(i)}  \Lambda^{(i-1)}  {A^{(i)}}^\dagger =  \Lambda^{(i)}, \; \forall \; 1\leq i \leq N \label{eq:pos},
    \end{eqnarray}
\end{subequations}
where $\Lambda^{(i)}$ are diagonal, completely (full-rank) positive, and unit trace preserving ($\text{Tr}[\Lambda^{(i)}]=1$) matrices. The \textit{isometry conditions} of Eq.~\ref{eq:iso} enable analytic partial trace while Eq.~\ref{eq:pos} preserves positivity under trace.

The computational complexity of evaluating expectation values, up to a controlled approximation error $\epsilon$, using the MPS scales then as $\mathcal{O}(\text{poly}  (N, \chi))$ operations. In the limit $\chi \ll 2^N$, where $N$ is the system size, the computational cost is low in comparison to the state-vector formalism. The opposite limit, of so called \textit{volume-law} states where $\chi$ is comparable with $2^N$, demands high computational resource cost and therefore volume-law systems with higher qubit numbers remain intractable using known classical methods. 

\section{MP$\rho$: A classical/quantum data-structure} \label{sec:MPρ}

\begin{figure}
    \centering
     \includegraphics[width=0.65\linewidth]{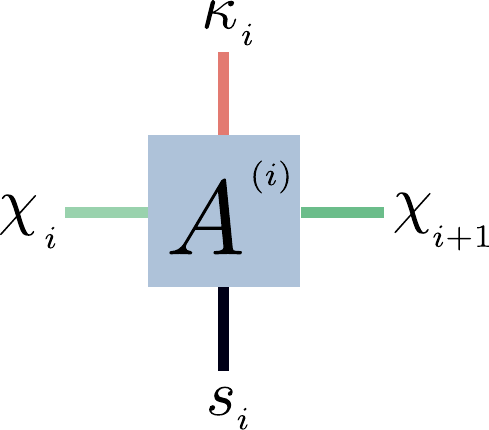}
    \caption{Generalization of the MPS tensor to that of the MP$\rho$ canonical form. We decorate each 
    tensor, $A^{i}$, equipped with a physical index, $s_{i}$ and virtual indices $\chi_{i}$ and $\chi_{i+1}$ 
    with an additional index $\kappa_{i}$.}
    \label{fig:forms}
\end{figure}

By decorating the MPS data structure, with additional virtual indices, one arrives at the MP$\rho$ data-structure. The decorated degree of freedom i.e., the new virtual index, denoted by the $\kappa$-index is illustrated in Fig.~\ref{fig:forms}, locally encodes mixture probabilities. This representation of incoherent mixtures, with a virtual index, is analogous to the role the $\chi$-dimensional bonds play in encoding non-local coherence correlations. We now briefly review the concept of mixture correlations. Afterwards, we discuss the holistic integration of coherence and mixture correlations into the MP$\rho$ canonical form. Lastly, we present generalized canonical constraints, analogous to the MPS's orthogonality center constraints, in the context of an MP$\rho$ canonical form.

We begin by noting a subtle difference between the Matrix Product (Density) Operator, MPO (MPDO)~\cite{Cuevas_2013, Verstraete_2004} and MP$\rho$. Namely, the virtual $\kappa$-mixture indices do not appear in the original MPO (MPDO) representation. As a result, one loses both manifest positivity as well as critical insight into the individual pure components comprising $\rho$. In contrast, the locally purified density operator (LPDO)\cite{Werner2016} representation does explicitly encode mixture correlations with an auxiliary $\kappa$ index. As such, LPDO is synonymous with MP$\rho$. For a more detailed list of similar representations, see Tab.~1 of Ref.~\cite{Ambroise24}. In order to further emphasize that a local purification is a natural extension of the 1D MPS (equipped with canonical constraints), we choose the use of the acronym MP$\rho$. 

A first theme of the current work, as is done when defining orthogonality centers in the MPS, is to provide operational constraints for the MP$\rho$ and highlight utility in gauging non-unique representations. As we will discuss in later sections, gauge invariances amongst the local purification representations will occur in the presence of global mixture correlations. Therefore, as another important point, we discuss how and when locally-purified data-structure efficiently encode \textit{global} mixture correlations. Practically, representing global mixture correlations involves selecting a representation which is suitable, in the sense that it minimizes the Kolmogorov-complexity of the representative data structure. One other key result of the current work is to provide a quick algorithm that integrates the two-body channel, within a completely positive MP$\rho$ framework without optimization routines. While there have been previous attempts at constructing such an algorithm, these methods rely on optimization routines~\cite{Werner2016, Ramirez_2024} or consider other memory intensive representations~\cite{Cheng_2021} thereby hindering the algorithm's overall scalability. 

\subsection{Classical Mixture correlations}
\label{sec:MCs}

The coherence and incoherent mixture correlations are both \textit{independently} normalized. The coherences are $L_2$-normalized, $\sum_i |c_i|^2 =1^2$, while the incoherent mixture correlations are $L_1$-normalized, $\sum_k p_k =1$. Note that the SVD is based on the 2-norm, and is therefore a natural fit for optimizing coherence correlations. When using SVD to optimize the $L_1$ normalized mixtures, an $L_2$-normalized interpretation is that $\sum_k \sqrt{p_k}^2 =1^2$. 

To see how both types of correlations co-exist, expand
\begin{eqnarray}\label{eq:rho_sum_representation}
    \rho &=& \sum_k p_k \rho_k = \sum_k p_k \ket{\psi_k}\bra{\psi_k} \nonumber \\
     &=& \sum_k p_k  \left[ \sum^{2^N-1}_{i=0}c^{(k)}_i \ket{i} \right] \bigotimes \left[ \sum^{2^N-1}_{j=0}c^{*(k)}_j \bra{j}\right] \\
     &=& \sum_{i,j,k} p_k \left[ c^{(k)}_i c^{*(k)}_j \ket{i}\bra{j} \right] \nonumber
\end{eqnarray} 
where $c^{(k)}$ are $\ket{\psi_k}$'s correlation coefficients and $\bigotimes$ denotes \textit{outer} product. Recalling $\bra{n}\ket{m} = \delta_{n,m}$, the trace then unifies both normalizations:
\begin{eqnarray}
    \text{Tr}\left[ \rho \right] &=& \sum_{l=0}^{2^N-1} \bra{l} \rho \ket{l} \nonumber \\ 
    &=& \sum_{k,i,j,l} p_k \left[ c^{(k)}_i c^{*(k)}_j \bra{l}\ket{i}\bra{j}\ket{l} \right] \nonumber \\
    &=& \sum_{k} p_k \sum_l |c^{(k)}_l|^2  = \sum_{k} p_k \times 1 \nonumber \\ &=& 1\times1=1! 
\end{eqnarray}

As introduced earlier, the dimension of the virtual $\kappa$-index provides insight into the mixture coefficients. That is, at a given site, dim($\vec{p}$) = $\kappa_i$. The intuition for MP$\rho$ is that the $k^{\text{th}}$ mixture coefficient $p_k$ encodes, via the $\kappa_i$ index, the probability for the positive operator $\vec{A}_k \vec{A}^\dagger_k$.   

Note that the MP$\rho$ formalism provides many possible non-unique gauge choices for factorizing the incoherent mixture correlations. For example, in  the \textit{local} mixture limit, we can absorb the \textit{global} $\vec{p}$ probabilities into single tensor i.e., by setting $p_k^{\frac{1}{2}} A^{(k)}_j=\tilde{A}^{(k)}_j$ for a single $j$. Alternatively, global mixtures could be symmetrically decomposed into all local $c_i$, simply by setting $p_k^{\frac{1}{2N}} A^{(j)} \rightarrow \tilde{A}^{(j)}_k$ for all $j$. Independent of the gauge one chooses to store correlations, a further non-uniqueness characteristic of MP$\rho$ is illustrated and detailed in  Sec.~\ref{sec:uni_freedom}.

In the rest of this section, we first review the canonical form before providing illustrative examples that develop an intuition regarding the functionality of the virtual $\kappa$ index. In doing so, we will provide examples in the opposite limits of local (see Sec.~\ref{sec:mprho_attract}) and global (see Sec.~\ref{sec:trace}) classical mixture correlations.

\subsection{MP$\rho$ canonical form}
\label{sec:MPR_CF}

The MP$\rho$'s non-negative mixture correlations are encoded within a completely positive matrix product operator (CPMPO)~\cite{Werner2016} representation. The key desirable feature of such a CPMPO is that it \textit{manifestly enforces positivity} via an explicit positive representation $\rho = \vec{A}^\dagger \vec{A}$, where $\vec{A} = \{ A_1, \cdots, A_N \}$ denotes the tensor train. As illustrated in Fig.~\ref{fig:forms}, the CPMPO's $A^{(i)}$ tensors are equipped with virtual $\chi$-dimensional coherence vector spaces, physical vector spaces $s_i$, and now additionally with $\kappa_i$-dimensional mixture spaces. This generalizes our prior definition for each site's tensor to
\begin{equation}
A^{(i)}:=A^{\chi_{i},s_{i},\kappa_{i}}_{\chi_{i+1}} = \mathds{V}_{\chi_{i}} \otimes \mathds{V}_{s_i} \otimes \mathds{V}_{\kappa_i} \otimes \mathds{V}^*_{\chi_{i+1}} \rightarrow \mathds{C}.
\label{eq:MPrho_tensor}
\end{equation}
Note that grouping $\chi_{i}$ and $\kappa_i$ with $s_i$ foreshadows a left-orthogonality decomposition. Indeed, grouping the top indices together preconditions the tensor for a $\text{top} \times \text{bot}$-dimensional matrix singular value decomposition. In the limit of $\kappa_{i}=1, \forall i$, one recovers the MPS data structure. Since it is not always clear how to factorize the mixture correlations, amongst the collection of individual tensors, the dimensionality of each tensors $\kappa$-index represents a powerful, latent gauge freedom that we investigate below. 

\subsection{Canonical Constraints}
\label{sec:MPR_CC}

The SVD of the tensor at the $i^\text{th}$ site reads 
\begin{eqnarray}
    A^{(i)} := A^{\chi_i, \kappa_i, s_i}_{\chi_{i+1}} = U^{\chi_i, \kappa_i, s_i}_{\sigma} \Sigma^\sigma_\sigma ( V^{\sigma}_{\chi_{i+1}} )^T
    \label{eq:rho_SVD}
\end{eqnarray}
where $\sigma =\text{dim}(\Sigma)=\text{min(dim}(U),\text{dim}(V^T)$). The indexing of the terms in Eq.~\ref{eq:rho_SVD} highlights a right orthogonalization, where the norm is absorbed into the neighboring tensor to the right ($i+1$). After this step, the updated tensors (denoted by overhead tilde) are given by
\begin{eqnarray}
     U^{\chi_i, \kappa_i, s_i}_{\sigma} &\rightarrow& \tilde{A}^{(i)}  \nonumber \\ 
     \sum_{\sigma, \chi_{i+1}} \Sigma^\sigma_\sigma ( V^{\sigma}_{\chi_{i+1}} )^T A^{\chi_{i+1}, \kappa_{i+1}, s_{i+1}}_{\chi_{i+2}} &\rightarrow& \tilde{A}^{(i+1)} 
\end{eqnarray}
where $\sigma$ is then re-labeled as the new $\chi_{i+1}$.

Having factorized a site tensor and subsequently absorbed the singular value core $\Sigma$ into the adjacent site, the updated tensor at site $i$ now obeys the canonical isometric condition. From the perspective of this first tensor, tracing out the first lattice site's physical ($s_i$), and mixture ($\kappa_i$), and left ($\chi_i$) degrees of freedom with its dual yields
\begin{eqnarray}
\label{eq:iso_rho}
\text{Tr}[A^{(i)}A^{(i)^\dagger}] := \sum_{\chi_i, \kappa_i, s_i} A^{\chi_i, \kappa_i, s_i}_{\chi_{i+1}} (A^{\chi_i, \kappa_i, s_i}_{\chi_{i+1}})^\dagger  = \mathds{1}_{\chi_{i+1}}
\end{eqnarray}
which is the isometric generalization of Eq.~\ref{eq:iso}. Like in the MPS case, given a judicious series of matrix decompositions and contractions absorbing the state's norm onto a single orthogonality core tensor, we can fulfill Eq.~\ref{eq:iso_rho} on all sites to the left and right of an OC.  

\subsection{Unitary freedom and Non-Uniqueness}
\label{sec:uni_freedom}
\begin{figure*}
    \centering
    \includegraphics[width=0.95\linewidth]{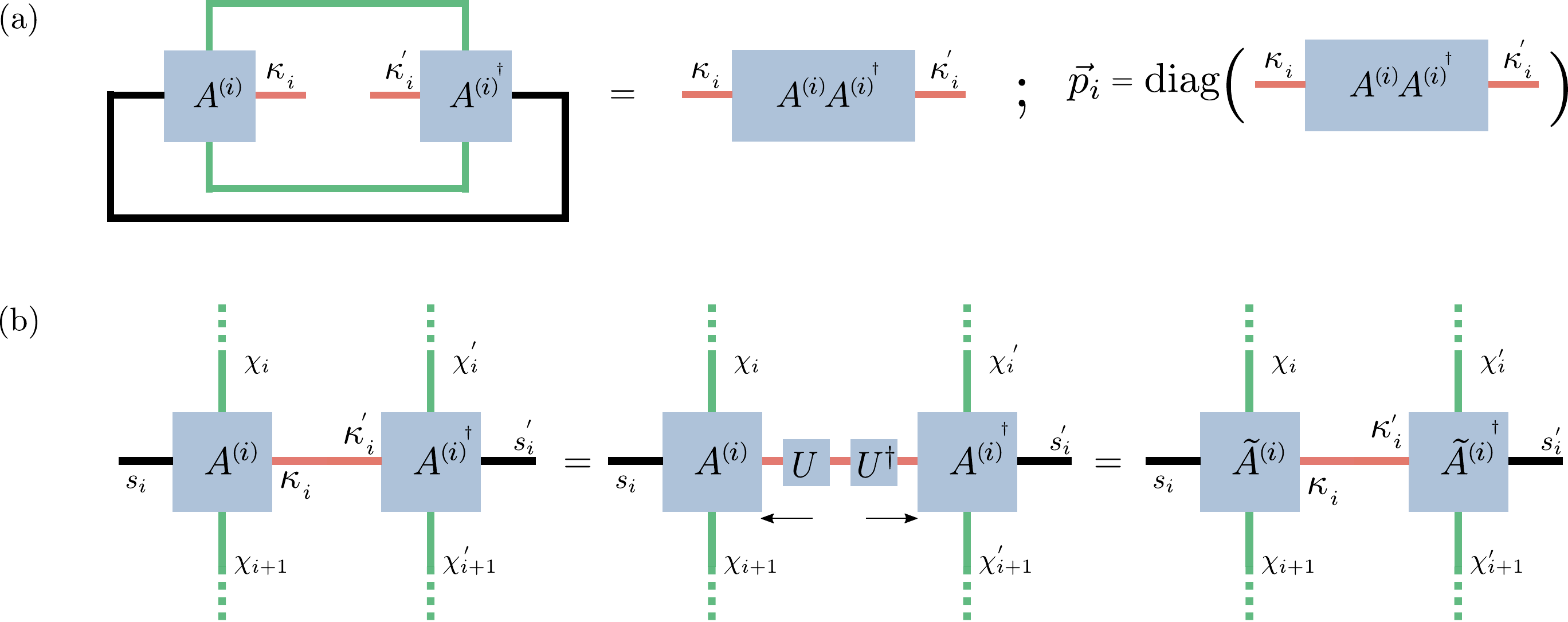}
    \caption{(a) Procedure to obtain the probabilities, $\vec{p}_{i}$ corresponding to the $\kappa$-subspaces at site, $s_i$. (b) Isometric gauge freedom of MP$\rho$ via an isometric map leaving the corresponding $\rho$ invariant. The application of such a map, in the $\kappa_i$ subspace, proves the non-uniqueness of the MP$\rho$ representation. This is directly analogous to the non-unique representation of MPS as illustrated in Fig.~\ref{fig:non_unique_mps}.}
    \label{fig:kappa_iso}
\end{figure*}

Earlier we described $\rho$, a mixed state, as a sum of positive operators $\vec{A}_{k}\vec{A}_{k}^{\dag}$ weighed by the corresponding mixture probability coefficients, $\vec{p}_{k}$, obtained as in Fig.~\ref{fig:kappa_iso}(a). In the following, we highlight the invariance of $\rho$ under the action of an internal transformation. That is, the action of isometries on $\vec{A}_{k}$'s $\kappa$-subspace, as illustrated in Fig.~\ref{fig:kappa_iso} (b) leaves the mixed state invariant. As a result, the mixture sum representation is non-unique and $\rho$ is represented up to an equivalence class of (most generally isometric) transformations of the $\kappa$-subspace. That is, 
\begin{eqnarray}
&  \rho = &\sum_{k}p_{k}A_{k}A_{k}^{\dag} \\ \nonumber
&       = & \sum_{k}p_{k}A_{k}U U^{\dag}A_{k}^{\dag} \\ \nonumber
&       = & \sum_{k'}p_{k'}\tilde{A}_{k'}\tilde{A}_{k'}^{\dag} = \rho'. \nonumber
\end{eqnarray}

\subsection{Gauge fixing two-body channels}
\label{sec:two_body_channel}
\begin{figure*}
    \centering
    \includegraphics[width=\linewidth]{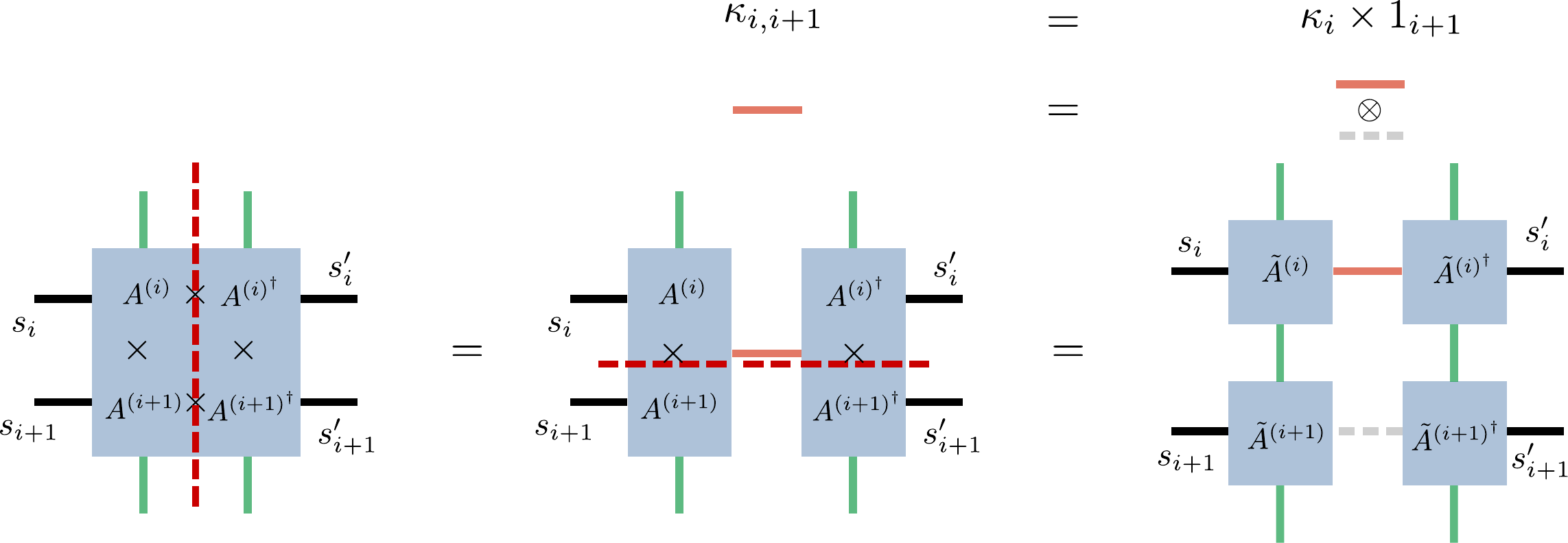}
    \caption{Optimization free routine for the application of the two-body noise channel.
    (Left) Factorization of $\rho_{i, i+1} = \sum\limits_{k=1}^{\kappa_{i, i+1}}A^{(i, i+1)}[:,k] A^{(i, i+1)}[:,k]^\dagger$. (Middle, Right) SVD performed across the site-wise bi-partition, by expressing $\kappa_{i, i+1} =  \kappa_i \otimes 1_{i+1}$, where $\kappa_i = \kappa_{i, i+1}$. The above decomposition results in a valid MP$\rho$ canonical form. The red dashed line represents decomposition using SVD.} 
    \label{fig:two_body_channel_op}
\end{figure*}

Before we delve into the numerical simulations in the next section, we present the operational procedure employed in applying a two-body channel operator. A key highlight of our approach is that the method introduced is optimization free, in comparison to other approaches wherein they rely on optimization routines in realizing the same~\cite{Werner2016}. The steps involved in realizing the operation are outlined in Tab.~\ref{alg:L2_application} and Fig.~\ref{fig:two_body_channel_op}. Mathematically, note the multiplicative gauge freedom in $\kappa_{i, i+1} = \kappa_{i} \otimes \mathds{1}_{i+1} =\mathds{1}_{i} \otimes \kappa_{i+1}$. That is, we set dim($\kappa_{i}$), after the gauge fixing, to be equal the prior dim($\kappa_{i,i+1}$) and then continue with the simulation algorithm. This is noteworthy because we retain the MP$\rho$ canonical form after the application of the two-body noise channel.

\begin{algorithm}[H]
    \begin{algorithmic}[1]
    \STATE Orthogonalize onto site $i$ or $i+1$,
    \STATE Trace over virtual index $\chi_{i}$, i.e., the virtual coherent bond shared between tensors at sites $s_{i}$ and $s_{i+1}$
    \STATE Multiply the two-body noise channel $\mathcal{E}_{i,i+1}(\cdot)$ and the two-body reduced density operator $\rho_{i,i+1}$ by contracting over all shared indices: $s_{i}, s_{i}', s_{i+1}, s_{i+1}', \chi_{i}, \kappa_{i}, \kappa_{i+1}$
    \STATE SVD the resulting $\tilde{\rho}_{i, i+1}$ in the transverse direction generating a combined mixture index $\kappa_{i, i+1}$ (Fig.~\ref{fig:two_body_channel_op} left panel)
    \STATE Combine $\chi_{i-1}, s_i, \kappa_{i,i+1}$ indices to SVD the resulting $A_i \times A_{i+1}$ tensor in the longitudinal direction (Fig.~\ref{fig:two_body_channel_op} middle panel)
    \STATE Decorate the tensor at site $i+1$ with a $\text{dim}(\kappa_{i+1}) = 1$ index (dashed grey line in Fig.~\ref{fig:two_body_channel_op} right panel)
    \RETURN Tensors $\tilde{A}_i, \tilde{A}_{i+1}$
    \end{algorithmic}
    \caption{Application of two-body noise channel}
    \label{alg:L2_application}
\end{algorithm}

\section{Numerical Results}
\label{sec:num_sim}
We now present a series of illustrative simulations that provide further insight into the inner workings and algorithmic applicability of MP$\rho$. To do so, we utilize 1- and 2-local Markovian quantum channels to transform entangled resource states into incoherent mixed states of varying locality. Afterwards, we validate the optimization-free two-body quantum channel algorithm by studying the noise's action on analytically solvable Greenberger–Horne–Zeilinger (GHZ) state dynamics. Finally, we examine the form of global mixture correlations that arise due to tracing out physical degrees of freedom, as in quantum erasure channels.  

\subsection{Local MP$\rho$ attractors}
\label{sec:mprho_attract}

\begin{figure}[t!]
    \centering
    \includegraphics[width=1\linewidth]{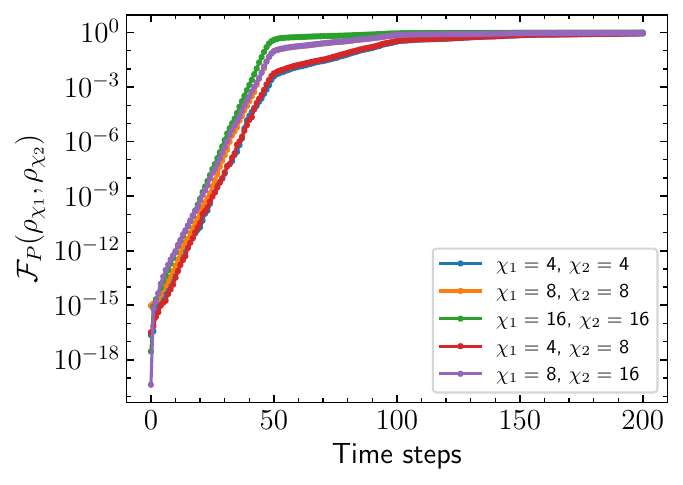}
    \caption{The flow of state fidelities as a function of time steps (clock-cycles) in the presence of 1-body bit and phase flip quantum noise channels. At $t=0$, we initialize random MPSs characterized by $\chi_1$ and $\chi_2$ (i.e., we use the \textit{randomMPS}
    functionality in ITensors that generates random MPSs whose bond dimensions are $\chi_1$ and $\chi_2$). These randomly initialized states have overlap fidelities $F_{ij} =\norm{\bra{\psi_i} \ket{\psi_j}}^2 \leq 10^{-14}$. By the end of the evolution in the presence of quantum channel the initially approximately-orthonormal states now have $10^{-1} \leq F_{ij} \leq 1$. The fidelity is computed at each time step with respect to a pair of evolved states, with the system size set to $N=50$. The total number of time steps is given by the number of cycles, $n_c$, times the number of sites. Each cycle has $N$ time steps and each time step has noise channels acting on a particular site. In the current scenario we set $n_c$=4. As the initial states are chosen randomly, the only features that are guaranteed to hold for any given pair, irrespective of the chosen $\chi_1$ and $\chi_2$ are: the orthonormality at $t=0$ and the flow to a completely depolarized state in the presence of noise.}
    \label{fig:MPpflow}
\end{figure}

To benchmark quantum computers, especially at scale, one may be interested in quantifying how 1-local dissipative maps destroy quantum coherence correlations. To examine this, we first prepare random MPS states which are characterized by different bond dimensions but defined on an identical $L$-site lattice. (See App.~\ref{sec:comparison} for details of how a $\log(\chi)$ depth random circuit can also produce such states.) Note that the product of the spatial extent with bond dimension quantifies the system's total entanglement. Next, we evolve the distinct random states under the action of 1-local noise channels. The noise operators act on the state at each time step, thus defining an open-quantum system clock cycle. 

Explicitly, the depolarizing quantum channel which is applied at each error clock cycle factorizes as the composition of two quantum maps. The first map is a single qubit phase ($p$) damping channel with Kraus operators $K^{p}_{0} = \sqrt{\alpha}\mathds{1}$ and $K^{p}_{1} = \sqrt{1-\alpha}Z_{i}$, where $Z_i$ is the Pauli-$Z$ acting on site $i$, with $\alpha$ being the dephasing rate. The second channel, the bit-flip ($b$) map, is characterized by the Kraus operators $K^{b}_{0} = \sqrt{\beta}\mathds{1}$ and $K^{b}_{1} = \sqrt{1-\beta}X_i$ with $X_i$ being the Pauli-$X$ acting on site $i$ and $\beta$ the bit-flip rate. 

Intuitively, for any almost-depolarized quantum state in the long time limit, we expect the action of the noise channels to drive the state towards a final form $\rho_{F} = (1-\epsilon)\mathds{1}/D + \epsilon\sigma$ where $\sigma$ is a non-local remnant of the initial entanglement. Note the important point that $\mathds{1}/D = \otimes_{j=1}^{N}\mathds{1}_{j}/2$ is generated by, and factorizes into, local identity operators. 

To quantitatively understand the noisy dynamics we compute a purity-based fidelity, $\mathcal{F}_P(\rho, \sigma)$. With $\ket{\Psi}, \ket{\Phi}$ as purifications of states $\rho$ and $\sigma$ respectively, the Uhlmann-Josza fidelity quantifies the closeness of states as
 \begin{equation}
 \label{eq:UJ}
     \mathcal{F}_{UJ}(\rho, \sigma) \defeq \max_{\{\ket{\Psi}, \ket{\Phi}\}}  |\ip{\Psi}{\Phi}|^2 = \text{Tr}[\sqrt{\sqrt{\rho} \sigma \sqrt{\rho}}]^2. 
 \end{equation}
Due to the difficulty of computing $\mathcal{F}_{UJ}$, the Hilbert-Schmidt inner product (HSIP), $\text{Tr}[\rho \sigma]$ has recently been investigated in terms of computing alternative fidelities which satisfy Josza's axioms~\cite{Liang2019}. Rather than its self-fidelity, HSIP yields the \textit{purity} of a state, $P(\rho) \defeq \text{Tr}[ \rho^2 ]\le 1$, with the equality holding only for pure states. However, the purity turns out to be a useful normalization function leading to the following alternative fidelity metric~\cite{Liang2019} 
 \begin{equation}
 \label{eq:UJ}
     \mathcal{F}_P(\rho, \sigma) = \frac{\text{Tr}[\rho \sigma]}{\max(P(\rho),P(\sigma))}.
 \end{equation}

The legend in Fig.~\ref{fig:MPpflow} indicates the fidelity is being taken with respect to random MPS with entanglement quantified by the bond dimension $\chi$. The initially distinct states, resulting from different noiseless quantum computations, then flow towards one another as evidenced by the fidelities flowing to unity. This illustrates how, after a short mixing time, the completely depolarized state is the fixed-point attractor of the local depolarizing Kraus dynamics. To perform the simulations, we program the MP$\rho$ type from the base types defined in the ITensors.jl~\cite{Matthew_2022} julia library. All the simulations are single core, single thread computations run on Intel Core Ultra 7 processor.

\subsection{GHZ stability under two-local channels}
\label{sec:GHZaction}

While methods for the application of single-body noise channels have been largely explored, methods for the two-local counterpart are more complicated. Namely, prior algorithms returning the states to the canonical form rely on optimization methods~\cite{Werner2016}. This optimization hinders the scalability of simulation protocols involving two-local channels. To this end, we employ the optimization free protocol introduced in Sec.~\ref{sec:two_body_channel}. In short, the mixture correlations are heuristically assigned to the $\kappa$ index of one of the two sites involved in the two-body channel. 

\begin{figure}[h!]
    \centering
    \includegraphics[width=\columnwidth]{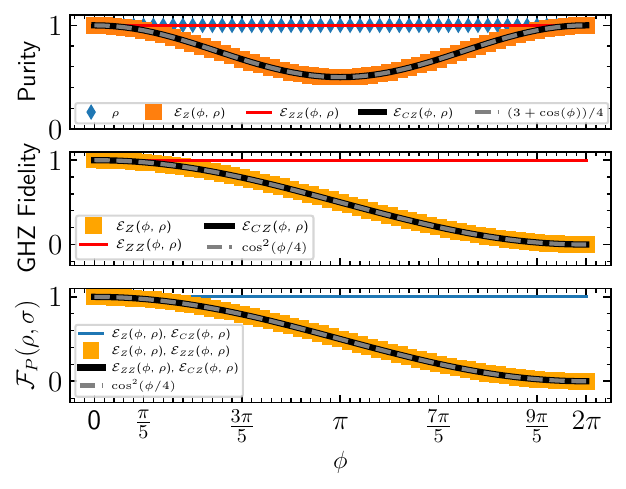}
    \caption{(Top) Purity of ideal GHZ (diamonds) versus that of a GHZ state corrupted by quantum channel described in terms of $Z_i, Z_i Z_{i+1}$, or $C_iZ_{i+1}$ (squares, red, and black lines respectively) Kraus operators. The bare GHZ state and $Z_iZ_{i+1}$ noised gates remain pure ($P=1$) while, the $Z_{i}$ and $C_iZ_{i+1}$ noised states have minimal purity $P=1/2$ at $\phi = 2 \arcsin(\sqrt{\alpha})=\pi$ where $\phi$ parameterizes the dissipative rates of the Kraus operators~\cite{Nielsen_2011}. (Middle) Fidelity between the ideal GHZ state and damped variants. The action of $Z_i Z_j$ stabilizes the state, while the $Z_{i}$ and $C_{i-1}Z_{i}$ gates also act identically and phase damp the GHZ state. (Bottom) Purity based fidelity of noised quantum states. Since $Z_i$ and $C_{i-1}Z_i$ dissipators act on the GHZ state identically, the fidelity of these two states is unity, but below unity when computing fidelity with respect to other states. We note that the results are computed for a system size of $N=500$, however these hold even in the thermodynamic limit of $N\rightarrow \infty$ i.e., they can be expressed analytically (dashed grey lines) with the analytical expressions derived as in App.~\ref{app:ghz_analytical}.}
    \label{fig:two_body_channel}
\end{figure}

To numerically validate our method, we apply distinct two-body noise channels on a $N$-qubit GHZ state: $\ket{\phi}=\frac{1}{\sqrt{2}}(\ket{0}^{\otimes N} + \ket{1}^{\otimes N})$. Denoting $\ket{\bm{i}} = \ket{i}^{\otimes N}$, the initial GHZ density matrix $ \rho_{GHZ} = 1/2(\ket{\bm{0}}\bra{\bm{0}} +\ket{\bm{0}} \bra{\bm{1}} + \ket{\bm{1}} \bra{\bm{0}}+ \ket{\bm{1}}\bra{\bm{1}})$. The analytic properties of the GHZ state, see App.~\ref{app:ghz_analytical}, provide a good platform to 
test and verify our results. 

We consider the action of quantum channels with three different types of dephasing noise. In the operator sum representation, these channels are described in terms of the Kraus operators: $K^{ZZ}_{0}=\mathds{1}$ and $K^{ZZ}_{1} = Z_{i}Z_{i+1}$, $K^{Z}_{0}=\mathds{1}$ and  $K^{Z}_{1} = Z_{i}$, and $K^{CZ}_{0}=\mathds{1}$ and $K^{CZ}_{1} = CZ_{i,j}$. Here $Z_{i}$ is the phase-flip as defined earlier, $CZ_{i,j}$ represents the controlled-$Z$ gate with $i(j)$ as the control(target) qubit. Note that, instead of decohering, the GHZ state is stabilized as an $+1$ eigenstate of both $K^{ZZ}_{0}$ and $K^{ZZ}_{1}$. Under the action of $CZ$, the $\ket{\bm{0}}$ component is not dephased while the $\ket{\bm{1}}$ is dephased, identically as under the action of $Z$. In Fig.~\ref{fig:two_body_channel} we compute the purity, fidelity with original GHZ state, and mutual fidelity $\mathcal{F}_P(\rho, \sigma)$ under the action of the above noisy channels and note the agreement with the analytic results, as discussed in App.~\ref{app:ghz_analytical}.

\subsection{Erasure as a source of global mixtures}
\label{sec:trace}

In this section, we analyze trace as an operation to generate \textit{global} mixture correlations. This is a limit which, to the best of our knowledge and despite its importance to experiments, has not been carefully examined within the LPDO formalism. By tracing out a sub-system, we will highlight both the ability to re-gauge global mixture correlations as well as recover the analytic isometric properties which are important defining features of the MPS and MP$\rho$. Since mixture correlations may be decomposed and structured according to their locality properties, the movement of global correlations is a limit worth examining due to their ubiquitous nature.

\begin{figure*}
    \centering
    \includegraphics[width=\linewidth]{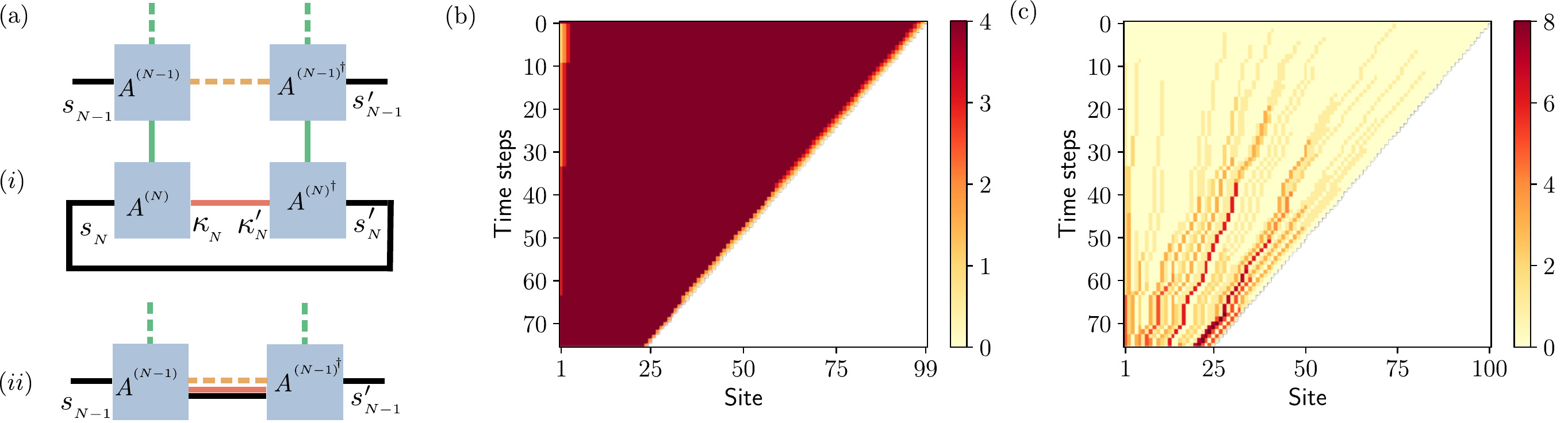}
    \caption{Tracing over local spin degrees of freedom generates mixture correlations as illustrated in panel (a). That is, tracing out site $s_{N}$ first involves (i) tracing over the physical indices, the corresponding $\kappa_{N}$ indices, and contracting over the virtual indices (the $\chi$-index) and then (ii) merging the local spin information into the adjacent mixture index $\kappa_{N-1}$ on the neighboring site. As a numerical illustration, we evolve a random state supported on $N=100$ sites and initially characterized by  $\chi_i$=16 and $\kappa_i$=1 $\forall i$. We then randomly choose a site to perform a partial trace and subsequently reorder the sites leading to (b) the translation of $\chi$-dimensional coherent correlations and to (c) growth in $\kappa$-dimensional mixture correlations as a function of the time. The colormap is on a $\log$-scale, with a maximum dimension of $16$ for $\chi$ and $256$ for $\kappa$.}
    \label{fig:ptrace_algo}
\end{figure*}

One of the simplest and paradigmatic physical examples where trace manifests is in single particle erasure\footnote{e.g. imagine an atomic trap where one atom is ejected from the trap. In this case one may not even know \textit{which} atom was lost, nullifying the notion of a third, flagged erasure state}. We model erasure of the $i^{\text{th}}$ qubit by a partial trace over the $i^{\text{th}}$ subsystem, leaving its $\bar{i}^{\text{th}}$ compliment behind. The operation as map on density matrix is given by $\rho \rightarrow \rho_{\bar{i}} = \text{Tr}_i \left[ \rho\right]$. The corresponding erasure channel in the MP$\rho$ formalism is to trace out the local site tensors. Explicitly, as illustrated in Fig.~\ref{fig:ptrace_algo}, this involves tracing over the physical degree of freedom, $s_{i}$ and the corresponding decorated mixture index, $\kappa_{i}$. Further, the corresponding latent virtual indices are contracted with an adjacent site, either $i-1$ or $i$. Tab.~\ref{alg:erasure_algo} outlines the algorithm for an erasure channel in the context of general quantum dynamics. In Fig.~\ref{fig:ptrace_algo}, we numerically simulate erasure, wherein at each time step we randomly trace out a site in a MP$\rho$ chain. The partial trace operation translates the coherent correlations into mixture correlations
evidenced by the growth in the $\kappa$-dimension post the trace.

\begin{algorithm}[H]
    \begin{algorithmic}[1]
    \STATE Select the $i^{\text{th}}$ qubit to erase. Reshape the tensor so that its physical indices and mixture indices are reshaped into a common index
    \STATE Contract the virtual indices (connecting with the $(i-1)^{\text{th}}$ or $(i+1)^{\text{th}}$ tensors)
    \STATE Re-orthogonalize the MP$\rho$, keeping the mixture indices with the orthogonality center, as desired
    \STATE Continue to apply one- and two-body unitaries or noise channels as the simulation requires
    \end{algorithmic}
    \caption{Erasure channel correlation conversion}
    \label{alg:erasure_algo}
\end{algorithm}

\subsubsection*{Gauging global correlations}
\label{sec:global_correlations}

In the following, we analyze the partial trace in the context of the GHZ state. The partial trace operation a qubit $i$ can be viewed as a map in the operator sum representation with Kraus operators given by $K_{0} = \bra{0}_{i}$ and $K_{1} = \bra{1}_{i}$. Partial trace of any spin results in the state $\mathcal{E}_{E}(\rho_{GHZ}) = 1/2(\ket{\bm{0}}\bra{\bm{0}}+ \ket{\bm{1}}\bra{\bm{1}})$, where the $\ket{\bm{0}}$ now represents tensor product over the remaining $N-1$ sites. As a result of partial trace, the off diagonal coherence correlations have been converted into diagonal mixture correlations. That is, we now have $p_0 = p_1 = \frac{1}{2}$ and both of these unentangled mixture components contain only \textit{one} coherence correlation ($c_{0,\cdots,0}=1$ \textit{or} $c_{1,\cdots,1}=1$ respectively) which has been renormalized to unity. 

\begin{figure*}[t]
    \centering
    \includegraphics[width=0.9\linewidth]{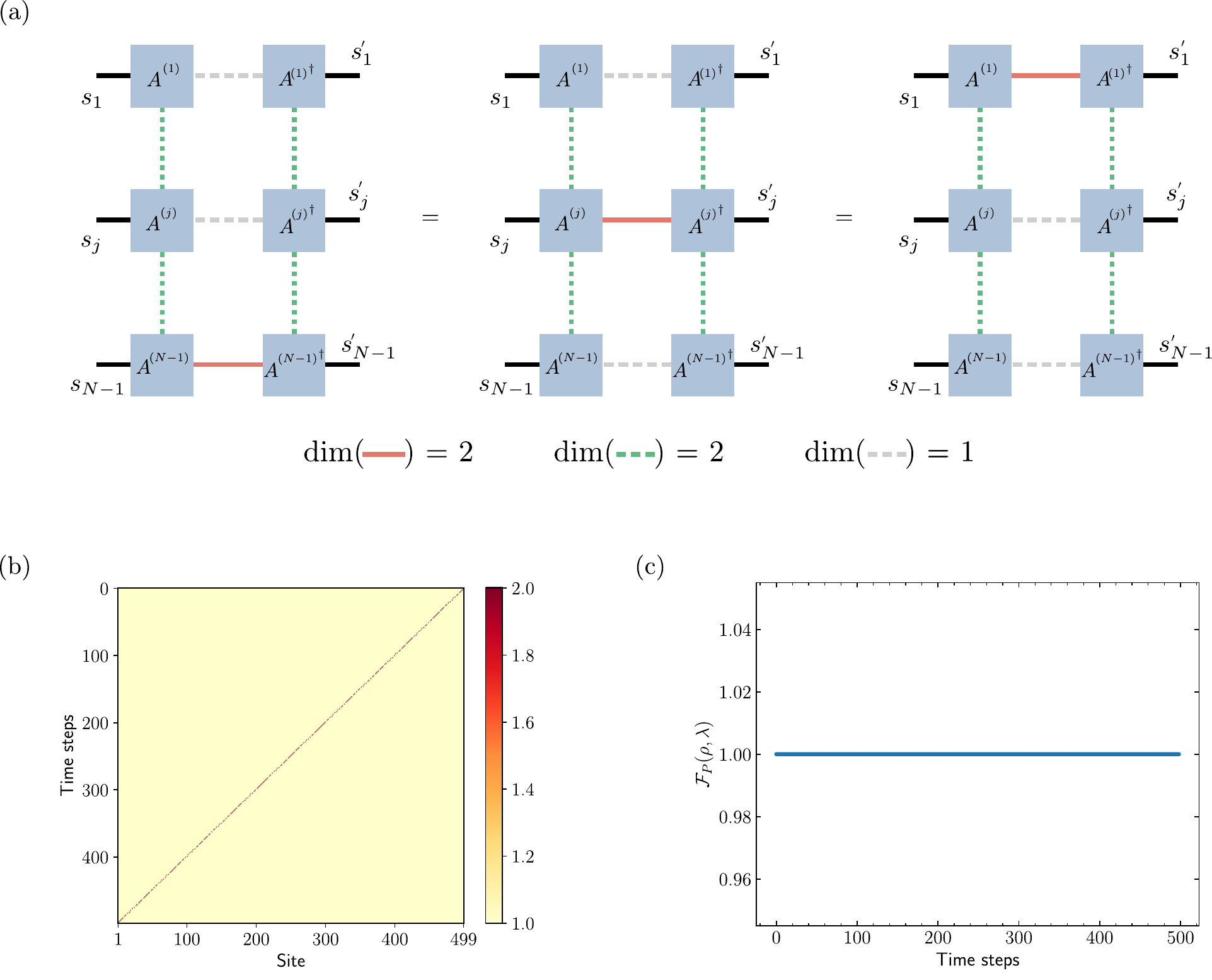}
    \caption{In the context of the GHZ state, after tracing out the $N^{\text{th}}$ qubit, the MP$\rho$ is internally re-gauged. 
    (a) This results in the freedom to store/transfer the mixture correlations at any site, as indicated by the 2-dimensional $\kappa$ index at the respective sites highlighted with a thick orange bond, such as site(left) $N-1$, (middle) some intermediate site $j$, or (right) the first site. We numerically validate the re-gauging procedure for a system of $N=500$ qubits. Here we evolve an initial state, obtained by tracing out the $N^{\text{th}}$ qubit, resulting in the $\kappa$-dimension at the $(N-1)^\text{th}$ site to be 2. The re-gauging protocol involves transport of the $\kappa$ to a neigbhoring site at each time step resulting in (b) the transfer of the $\kappa$-dimension. (c) The fidelity, $\mathcal{F}_{P}(\rho, \lambda)$, between the initial state $\rho$ and the re-gauged states, $\lambda$, at each step remains unity reflecting the invariance of $\rho$ under the $\kappa$-transport.}
    \label{fig:gauging_global_correlations}
\end{figure*}

The partial trace above highlights a few important but distinct invariances. The first being that we could have traced \textit{any} qubit, with the result still holding with the same global correlations. This is a consequence of the permutation invariance of the GHZ state. Secondly, we may, after trace, translate the location of the dimension-2 $\kappa$-index to any site of choice as highlighted in Fig.~\ref{fig:gauging_global_correlations}.

Consequently, we see how MP$\rho$'s latent invariances, in the possible representations one can gauge, enable some freedom in representing the mixture correlations. In this example, mixture correlations can be translated in the vertical direction as in Fig.~\ref{fig:gauging_global_correlations}. This is achieved by contracting nearest neighbor tensors and then setting the dimensions of the desired SVD as per the two-body optimization free update (Sec.~\ref{sec:two_body_channel}). Explicitly, in the case of a traced out qubit of the GHZ state, one sets the kappa index from 2 to 1, on the tensor currently with the index, while setting the kappa index from 1 to 2, on the tensor which will encode the mixture correlations after the next update. In Fig.~\ref{fig:gauging_global_correlations}, we elucidate the above using a numerical example. To this extent, we choose a $N=500$ GHZ state and perform a partial trace on the last spin, leading to the state, say $\rho$, the $\kappa$ dimension of which 
at the site $N-1$ turns 2 due to the trace. We then evolve the state, $\rho$, wherein at each time step we transport the $\kappa$ from one site to its left neighbor, resulting in the new state $\lambda$ at each time step until we reach the first site, thereby sweeping across the entire length of the chain. At each time step we compute the fidelity between $\rho$ and $\lambda$ and show that it indeed is unity, thereby establishing the invariance of $\rho$ under the $\kappa$-transfer.

\subsubsection*{Unitary freedom and Non-Uniqueness}

In the following we discuss the unitary freedom and non-uniquess of MP$\rho$ in the context of the GHZ state. We begin by tracing out a site at the end of the chain and denote $\ket{0}_N, \ket{1}_N$ to be the computational basis states of the traced out qubit. Each of them was initially coherently correlated with the rest of the spin-chain. Note that all entanglement properties of the system are invariant under local $SU(2)$ transformations. Additionally note that observable quantities are defined with respect to the basis independent trace $\langle \hat{A} \rangle = \text{Tr}[\hat{A} \rho]$. Fig.~\ref{fig:kappa_iso} (b) illustrates the freedom to unitarily transform a tensor across the local virtual mixture index. In the present case, of tracing 
a single qubit from a multi-qubit GHZ, the unitary freedom on the two Feynmann paths
is realized by interchanging the $\kappa$-indices in the gauge fixing protocol as in Fig.~\ref{fig:two_body_channel_op}. The invariance is proved by recalling that all observables are independent of the choice of basis when tracing out the qubit.


\section{Conclusion} \label{sec:conclusions}

In this work we have examined the operational interpretation of MP$\rho$ as a classical-quantum data structure that encodes both complex coherence and real mixture correlations. Building on prior LPDO efforts, our work develops technical algorithms to update the canonical form while also developing a physical and operational interpretation for the degrees of freedom encoded by the virtual mixture indices. In short, this investigation expands the computational tool-set for working with density operators at scale. 

Within the MP$\rho$ formalism we have first examined the evolution of entangled quantum states under the action of one- and two-local quantum channels. Previously, the action of two-local channel operators was defined in terms of a SVD in conjunction with a bond-dimension-minimizing optimization routine~\cite{Werner2016}. To simplify the application of two-local noise channels, we develop an optimization-free update which returns the tensors, contracted under the application of a two-body channel, back to the canonical form. To verify our technique, we have tested the application of two-body noise channels within the context of an analytically solvable example. Lastly, we present an algorithm that elucidates partial trace in the MP$\rho$ formalism and highlight the representation of global mixture correlations within a local gauge.

To stress test the computational capabilities of this framework, we present several numerical simulations involving single and two-body channel operators. For example, we have dissipatively evolved initial random $\chi$-entangled 50-qubit density operators. In contrast, naive simulations of density operators within a state-vector format are limited to systems with fewer than 25 qubits, even at supercomputing scales. Our work, therefore, highlights the large locus of mixed quantum states which are amenable to a classical MLRep and subsequently simulation.

The MP$\rho$ formalism provides tools to explore avenues that have been recently gaining traction~\cite{kato_2024, Srinivasan_2021}. For example, analyzing the dynamics of many-body systems in an open setting i.e., in the presence of the noise, can be hugely accelerated~\cite{Shah_2024,Sun_2024,Chen_2024,Wanisch_2025}. The simulation of quantum systems necessitating supercomputer resources, thereby strongly demarcating the classical simulable limits, remains an intriguing open question~\cite{Chen_2023}. We conjecture that, in future work, sampling of the decorated mixture index will be useful in parallelized algorithms which themselves might provide crucial insights into the fundamental limits of simulating open quantum systems. Furthermore, purified density operator MLReps can be constructed based on other tensor network canonical forms.  For instance, tree tensor networks~\cite{PhysRevA.96.062322, Arceci_2022, Ballarin2025, dubey2025simulating}, based on a heirarchical data-structure, are a promising candidate for further  exploration of a wider variety of noisy quantum systems~\cite{Sulz_2023, Li_2024}.

From an experimental viewpoint, MP$\rho$ tools can be deployed to not only characterize, but also subsequently mitigate, errors in current quantum architectures, thereby proving a useful tool for future quantum control theory~\cite{Mangini_2024}. A last future grand challenge is to analyze quantum algorithms, in terms of their computational complexity, in the presence of realistic noise models and, if possible, to optimize the associated performance metrics.

During the writing of this manuscript Refs.~\citenum{Ambroise24, kato_2024} appeared. These works also target simulations of noisy quantum systems, in terms of positive matrix product operators, but differ from our algorithm and results. 

\section{Acknowledgements}

E.D. and A.J. are supported by the U.S. Department of Energy, Office of Science, Advanced Scientific Research Program, Early Career Award under contract number ERKJ420. We thank A. Nocera, G. Alvarez, B. Xiao, V. Protopopescu, and ORNL's Quantum Information Science Section for valuable discussions. 

\appendix
\section{The coherence correlation tensor}
\label{app:C_tensor}

\begin{figure*}
 \centering
    \includegraphics[width=0.9\textwidth]{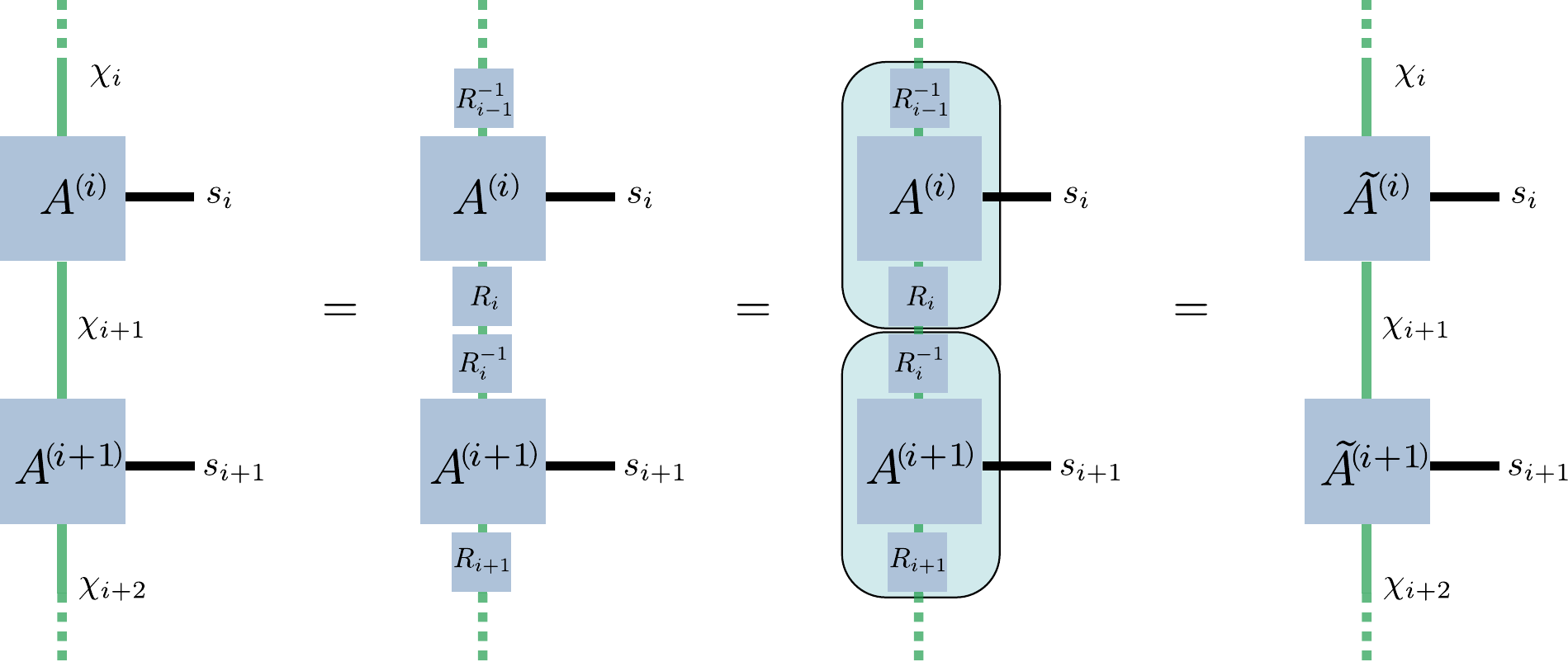}
    \caption{Non-uniqueness of MPS-representation as a mathematical gauge invariance. Injecting $\chi_i$-dimensional matrix resolutions of identity $R_i R^{-1}_i  = \mathds{1}_{\chi_i}$ into the first column's virtual correlation spaces results in the second column. Contracting the third column's shaded regions results in the fourth column's equivalent representation.}
  \label{fig:non_unique_mps}
\end{figure*} 

In this appendix, we briefly present the different approaches involved in representing the coherent correlations, $\bm{c}$. One is 
a global perspective, wherein $\bm{c}$ is vectorized into a $1$-tensor by grouping all the indices together into a global $2^N$-dimensional index such 
that $\bm{c} := \mathbb{C}^{2^N} \rightarrow \mathbb{C}$.  In other words, for given a computational basis vector, a complex amplitude 
associated to the bit string is returned, as $\bm{c} := \mathbb{C}^2 \otimes \mathbb{C}^2 \otimes \cdots \otimes \mathbb{C}^2 \rightarrow \mathbb{C}$. 
Given this \textit{local} tensor product structure $\bm{c}$ is an $N$-linear tensor, where each index is two-dimensional. 

MLReps of $\bm{c}$ go beyond this by encoding the complex correlations as a function of additional virtual indices. This is somewhat counter-intuitive as virtual indices are not a necessary condition and seemingly add complexity. However, as discussed in the main text, they provide a systematic way to express a locus of states containing only polynomial numbers of non-zero correlations in a scalable manner. Explicitly Eq.~\ref{eq:MPS} uses
\begin{equation}
     \bm{c} = \text{Tr}_\chi[A^{(1)} \cdots A^{(N)}].
\end{equation}

By inserting and re-contracting invertible matrices $R_i$ and $R^{-1}_i$ at the $i^\text{th}$ bond, Fig.~\ref{fig:non_unique_mps} graphically proves the non-uniqueness of a 1D tensor train. By judiciously leveraging this freedom, we can write the MPS in a unique left/right/mixed canonical form using and enforcing Eq.~\ref{eq:iso} and Eq.~\ref{eq:pos}. For further details see Ref.~\citenum{SCHOLLWOCK2011}.

\section{U(1) gauge-invariance}
As a gauge invariance warm up, consider a global $U(1)$ transformation  $|\psi_k \rangle \rightarrow e^{i\phi} |\psi_k \rangle$. This transforms the statistical ensemble as: $\rho \rightarrow  \sum_k p_k e^{i\phi}|\psi_k \rangle \langle \psi_k|e^{-i\phi} = \rho$. Since each of the $p_k$'s is positive we may write them in the complex polar form as $p_k = |p_k|e^{\text{Arg}({p_k})}=|p_k|$. This latent gauge transformation ensures that the mixture coefficients remain positive, non-complex quantities. That is, we can absorb these phases into the mixture \textit{eigenvalues} with opposing phases $\sqrt{p_k} e^{i\phi}\times \sqrt{p_k}e^{-i\phi}=p_k$. This phase cancellation thus provide an intuition for how other latent, locally invariant, gauged degrees of freedom work leading to the construction of equivalent $|\Psi\rangle$- and $\rho$-representations.

\section{The operator sum representation}

Let us examine some properties of quantum channels and their impact on mixture and coherence correlations. In the operator sum representation we expand a quantum channel as a sum over a basis of $L$ Krauss operators as 
\begin{equation}
     \label{eq:Krauss}
     \mathcal{E}(\rho) = \sum_{k=1}^{k=L}F_k \rho F^\dagger_k. 
\end{equation}
In order to separate the coherence correlations and the mixture correlations, so that we can ideally represent them with additive scaling, let us first examine the action of unitaries and quantum channels on \textit{coherence} (not mixture) correlations.

First, it is well known that the norm of the state's coherence correlations is unitarily invariant. That is, if $\ket{\Phi} = U \ket{\Psi}$, then $\sum_i |c^{(\Phi)}_i|^2 = \sum_i |c^{(\Psi)}_i|^2 = 1$. In either case, the norm of the coherences is manifestly preserved and the \textit{mixture} correlations are also invariant under such an operation since,
\begin{eqnarray} 
\label{eq:rho_invariance} 
U \rho U^\dagger&=&\sum_{k=1}^{\kappa} p_k U|\psi_k \rangle \langle \psi_k| U^\dagger \nonumber \\
                &=&\sum_{k=1}^{\kappa} p_k |\phi_k \rangle \langle \phi_k|.
\end{eqnarray} 
Again, the mixture correlations are $\vec{p}$ in both cases, with only difference being that the coherence correlations have been updated, but with their Euclidean norm preserved. In this special case, of a unitary quantum channel with $L=1$, the number of mixture correlations is \textit{non-increasing}.  

On the other hand, a quantum channel described by $k>1$ Kraus operators may modify both the coherence and the mixture correlations. As a general rule, a quantum channel with $L$ Kraus operators (in its minimal description) decreases the number of coherence correlations if it consists of one-body operators. Depending on the form of both the input state and the quantum channel the number of mixture correlations may increase from $\kappa$ to a maximum of $L\kappa$ or the number of mixture correlations may decrease to the minimum of $1$. 

\section{Comparison to other simulators}
\label{sec:comparison}
In this section we benchmark the results obtained using MP$\rho$ formalism to those obtained using QuTiP~\cite{qutip_2024}. To this extent, we consider a variant of the random quantum circuits as outlined in Refs.~\cite{Arute_2019,Huang_2020} in the presence of noise. That is, the circuit is made up of $d$ layers with each layer consisting of a single qubit unitaries randomly chosen from the gate set $\{\sqrt{X}, \sqrt{Y}, \sqrt{W}\}$, as in Eq.~\ref{eq:r_mat_1} and \ref{eq:r_mat_2} below, followed by the application of a two-qubit unitary $\text{fSim}(\theta, \varphi)$, parameterized by $(\theta, \varphi)$, as in Eq.~\ref{eq:r_mat_3}
\begin{align}
\sqrt{X} =\frac{1}{\sqrt{2}}\begin{bmatrix}
1 & -i \\ 
-i & 1    
\end{bmatrix},
\sqrt{Y} =\frac{1}{\sqrt{2}}\begin{bmatrix}
1 & -1 \\ 
1 & 1    
\end{bmatrix},
\label{eq:r_mat_1}
\end{align}

\begin{align}
\sqrt{W}=\begin{bmatrix}
1 & -\sqrt{i} \\ 
\sqrt{-i} & 1    
\end{bmatrix},
\label{eq:r_mat_2}
\end{align}

\begin{align}
\text{fSim}(\theta, \varphi)=\begin{bmatrix}
1 & 0 & 0 & 0 \\ 
0 & \cos(\theta) & -i\sin(\theta) & 0 \\
0 & -i\sin(\theta) & \cos(\theta) & 0 \\
0 & 0 & 0 & e^{-i\varphi}
\end{bmatrix}.
\label{eq:r_mat_3}
\end{align}

\begin{figure*}
 \centering
    \includegraphics[width=\textwidth]{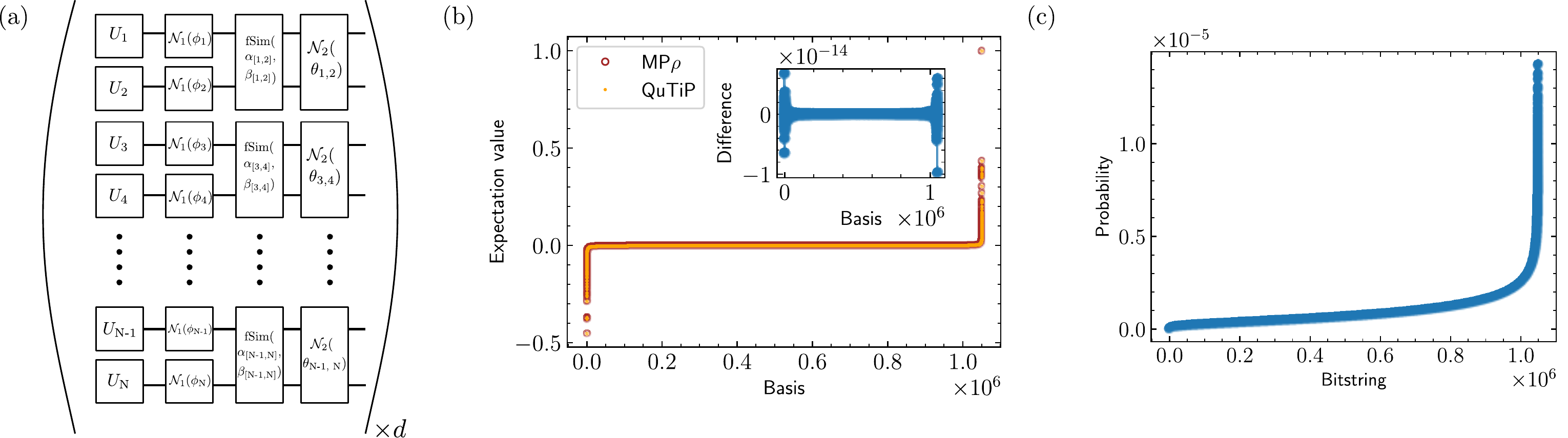}
    \caption{Cross validation of results obtained using MP$\rho$ formalism and QuTiP. (a) $d$-layer random quantum circuit consiting of single and two-qubit unitaries being appended with noise channels. The two qubit unitaries are parameterized by $(\alpha, \beta)$ and are chosen randomly from a flat distribution. The one-body noise channels at site $i$ are parameterized by $\phi_{i}$ while the two-body noise channels at sites, $(i, i+1)$ are parameterized by $\theta_{i, i+1}$. (b) The Pauli tomography of the final density matrix obtained after the noisy evolution of a of $10 \times 10$ ($ N \times d$) random quantum circuit using MP$\rho$ and QuTiP, and (inset) their difference showing agreement up to machine precision. For better representability, we have sorted the basis in the ascending order of their expectation values. (c) The probability of $2^{20}$ basis bitstrings with respect to the final density matrix of a $20 \times 20$ ($N \times d$) random quantum circuit with the basis sorted in the ascending order of their probabilities. The results are computed using the MP$\rho$ formalism whereas QuTiP requires more memory resources for executing the same $20 \times 20$ random quantum circuit instance. We notice a small number of bitstrings deviating from the flat distribution, as in the completely depolarized state, corresponding to a quasi-concentrated Porter-Thomas distribution. This behavior is as expected in the experimental realizations of random quantum circuits that are inherently noisy.}
  \label{fig:rand_cir}
\end{figure*}

Further, we apply noise by appending noise channels for the single and two-qubit unitaries, see Fig.~\ref{fig:rand_cir}. To this extent, single qubit unitaries in odd (even) layers are followed by dephasing (bitflip) noise while the two qubit unitaries are followed by $C_{i}Z_{i+1}$ noise channel for two layers and then by $Z_{i}Z_{i+1}$ for the subsequent next two layers, thereby repeating the pattern every four layers. Further to randomize, we randomly choose the noise rates for each of the noise channels parameterized by $\phi$ as detailed in App.~\ref{app:ghz_analytical}. To cross validate the results, we consider a $10 \times 10$ ($N \times d$) square circuit and compute the tomography of the final density matrix in the Pauli basis. As noted by Fig.~\ref{fig:rand_cir}, the results agree up to machine precision. We also compute the bitstring probabilities with respect to the final density matrix, $\bra{\psi}\rho\ket{\psi}$, of a $20 \times 20$ square circuit and note that a similar computation on QuTiP remains prohibitive, as the noisy circuit update demands higher memory resources than used by the MP$\rho$ formalism. Note that the final bitstring probabilities reflect a strongly depolarized Porter-Thomas distribution. 

\section{GHZ states under noise}
\label{app:ghz_analytical}
In this section we detail the action of the single body and two body noise channels on the GHZ state and their related properties. We begin by reviewing the Kraus operator representation of the different noise channels. The action of noise channel on a density matrix, $\rho_{s}$, can be summarized by 
\begin{equation*}
\rho' = \Tr_{\text{env}}[U (\ket{e_{0}}_{\text{env}}\bra{e_{0}} \otimes \rho_{s}) U^{\dag}]
\end{equation*}
where the system is coupled with an environment ancilla qubit initialized in $\ket{e_{0}}$ state. The above can be equivalently represented as 
\begin{eqnarray*}
\rho' &=& \sum_{k}E_{k}\rho_{s}E_{k}^{\dag}, \\
E_{k} &=&(\bra{e_{k}} \otimes \mathds{I}_{s})U(\ket{e_{0}} \otimes \mathds{I}_{s}) 
\label{eq:kraus}
\end{eqnarray*}
with $\ket{e_{i}}$'s representing the orthogonal basis that define the environment.

The choice of $U$ allows for the realization of different noise channels. Single qubit dephasing noise channel
can be realized by setting $U = \mathds{1}_{\text{env}} \otimes \ket{0}_{s}\bra{0} + R_{Y}(\phi)_{\text{env}} \otimes 
\ket{1}_{s}\bra{1}$ resulting in the Kraus operators
\begin{eqnarray*}
E^{Z}_{0} &= &\ket{0}_{s}\bra{0} + \text{cos}(\phi/2)\ket{1}_{s}\bra{1}, \\
E^{Z}_{1} &= &\text{sin}(\phi/2)\ket{1}_{s}\bra{1}. 
\end{eqnarray*}
Note that the Kraus operators $E^{Z}_{0}$, $E^{Z}_{1}$ are equivalent to the Kraus operators $K^{Z}_{0}$, $K^{Z}_{1}$ introduced in the main text via the unitary freedom of the operator sum representation (see exercise 8.27 in Ref.~\citenum{Nielsen_2011}). Likewise, the two qubit noise channels represented by the Kraus operators $K^{ZZ}_{0}$ and $K^{ZZ}_{1}$ can be mapped to $E^{ZZ}_{0}$ and $E^{ZZ}_{1}$ given by
\begin{eqnarray*}
E^{ZZ}_{0} &=& \ket{00}_{s}\bra{00} + \text{cos}(\phi/2)\ket{01}_{s}\bra{01} \\
& + & \text{cos}(\phi/2)\ket{10}_{s}\bra{10} + \ket{11}_{s}\bra{11} \\
E^{ZZ}_{1} &=& \text{sin}(\phi/2)\ket{01}_{s}\bra{01} + \text{sin}(\phi)\ket{10}_{s}\bra{10}
\end{eqnarray*}
obtained by setting 
\begin{eqnarray*}
U & = &\mathds{I}_{\text{env}} \otimes \ket{00}_{s}\bra{00} + R_{Y}(\phi)_{\text{env}} \otimes \ket{01}_{s}\bra{01} \\
  & + & R_{Y}(\phi)_{\text{env}} \otimes \ket{10}_{s}\bra{10} + \mathds{I}_{\text{env}} \otimes \ket{11}_{s}\bra{11}
\end{eqnarray*}

Similarly, the other two qubit noise channel introduced in the main text represented by the Kraus operators $K^{CZ}_{0}$ and $K^{CZ}_{1}$ can be equivalently represented by $E^{CZ}_{0}$ and $E^{CZ}_{1}$ given by
\begin{eqnarray*}
E^{CZ}_{0} &=& \ket{00}_{s}\bra{00} + \ket{01}_{s}\bra{01} + \ket{10}_{s}\bra{10}\\
& + &  \text{cos}(\phi/2)\ket{11}_{s}\bra{11} \\
E^{CZ}_{1} &=& \text{sin}(\phi/2)\ket{11}_{s}\bra{11}
\end{eqnarray*}
obtained by setting 
\begin{eqnarray*}
U & = &\mathds{I}_{\text{env}} \otimes \ket{00}_{s}\bra{00} + \mathds{I}_{\text{env}} \otimes \ket{01}_{s}\bra{01} \\
  & + & \mathds{I}_{\text{env}} \otimes \ket{10}_{s}\bra{10} + R_{Y}(\phi)_{\text{env}} \otimes \ket{11}_{s}\bra{11}
\end{eqnarray*}

Having obtained the Kraus operators for the different noise channels, in the following we present the action of the same on a $N$-qubit GHZ state, given by $\rho_{GHZ}$ as in the main text. We denote the final density matrix obtained by the action of single qubit dephasing channel, two-qubit $ZZ$ noise channel and two-qubit $CZ$ noise channel by $\rho^{DP}_{GHZ}$, $\rho^{ZZ}_{GHZ}$, $\rho^{CZ}_{GHZ}$ respectively and note that the final density matrices are given by
\begin{eqnarray*}
    \rho^{DP}_{GHZ} & = \rho^{CZ}_{GHZ} = & \frac{1}{2}\bigg(\ket{\bf{0}}\bra{\bf{0}} + \text{cos}(\phi/2)\ket{\bf{0}}\bra{\bf{1}}) \\
                    & + & \text{cos}(\phi/2)\ket{\bf{1}}\bra{\bf{0}} + \ket{\bf{1}}\bra{\bf{1}}\bigg)\\
    & \rho^{ZZ}_{GHZ}  = & \frac{1}{2}\big(\ket{\bf{0}}\bra{\bf{0}} + \ket{\bf{0}}\bra{\bf{1}}) 
                      + \ket{\bf{1}}\bra{\bf{0}} + \ket{\bf{1}}\bra{\bf{1}}\big).
\end{eqnarray*} 

We note that under the action of the two qubit $ZZ$ noise channel the GHZ state remains invariant. While the purity of the final density matrix, given by $\Tr[\rho^2]$ is one in the case of the GHZ state remaining invariant, in the other case of the single-qubit dephasing and the the two-qubit $CZ$ noise it is given by $(3 + \cos(\phi))/4$. The HSIP fidelity of $\rho^{DP}_{GHZ}$ and $\rho^{ZZ}_{GHZ}$ is given by $\cos^2(\phi/4)$. The measures involving other combinations can be evaluated in a similar fashion from the corresponding density matrices, as detailed above.

\section{Time profiling of noise channel application}
\begin{figure}[t!]
 \centering
    \includegraphics[width=\columnwidth]{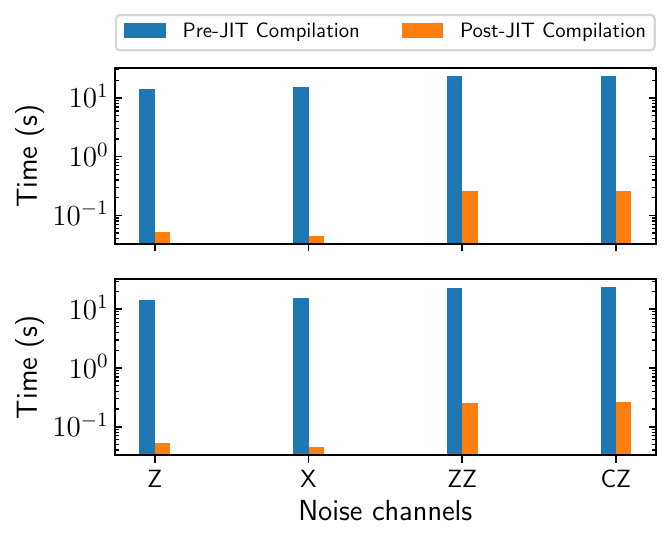}
    \caption{Profiling the performance of the application of one-body dephasing and bitflip noise channels via the Kraus operators $\{K_{0}^{p}, K_{1}^{p}\}$ and $\{K_{0}^{b}, K_{1}^{b}\}$ which are described in Sec.~\ref{sec:mprho_attract}. The two-body noise channels ZZ and CZ
    can be expanded in terms of the Kraus operators $\{K_{0}^{ZZ}, K_{1}^{ZZ}\}$ and $\{K_{0}^{CZ}, K_{1}^{CZ}\}$ respectively. The time required for the single application of the above noise channels on (Top) 50-qubit random MPS, (Bottom) 500-qubit GHZ state. The JIT compilation boosts the performance by atleast $\approx91$ times (in the worst case scenario) in comparison to the pre-JIT compilation times.}
  \label{fig:app_time}
\end{figure}

In this section, we present the time profiling of the application of one and two-body noise channels. To do so, we compute the wall-clock time, see Fig.~\ref{fig:app_time} required for the application of the one and two-body noise channels on a 50-qubit random MPS and on the 500-qubit GHZ state. We profile the time performance of applying four different noise channels which are considered in the main text. The pair of single-qubit dephasing and bitflip noise channels,  described in terms of the Kraus operators $\{K_{0}^{p}, K_{1}^{p}\}$ and 
$\{K_{0}^{b}, K_{1}^{b}\}$ respectively as described in Sec.~\ref{sec:mprho_attract}. Further, as described in Sec.~\ref{sec:GHZaction}, we also consider the two two-qubit noise channels described by the Kraus operators 
$\{K_{0}^{ZZ}, K_{1}^{ZZ}\}$ and $\{K_{0}^{CZ}, K_{1}^{CZ}\}$. We note that Julia offers a just-in-time (JIT) compiler. However, once JIT compiled, the function signature can be re-used and therefore we present the profiling pre and post-JIT compilation.

\bibliography{main}
\end{document}